\documentclass[conference]{IEEEtran}
\usepackage{graphicx}
\usepackage{caption}
\usepackage{subcaption}

\usepackage{longtable}
\usepackage{stfloats}
\usepackage{graphicx}

\makeatletter

\usepackage{ifpdf}
\usepackage[T1]{fontenc}
\usepackage{amssymb}
\usepackage{cite}

\ifCLASSINFOpdf
   \usepackage{graphicx}
	\usepackage{epstopdf}
   \graphicspath{{images/}}   
   \graphicspath{{../pdf/}{../jpeg/}{../eps/}}
   \DeclareGraphicsExtensions{.pdf,.jpeg,.png,.eps}
\else
   \usepackage[dvipdfm]{graphicx}
   \usepackage{bmpsize}
   \graphicspath{{../eps/}}
   \DeclareGraphicsExtensions{.eps}
\fi
\graphicspath{{images/}}

\usepackage[cmex10]{amsmath}
\usepackage{multirow}

\usepackage{multirow}

\begin{document}

\title{Machine Learning Based Obstacle Detection for Automatic Train Pairing}
%
\author{\IEEEauthorblockN{Raja Sattiraju, Jacob Kochems and Hans D. Schotten}
\IEEEauthorblockA{Chair for Wireless Communication \& Navigation \\
University of Kaiserslautern\\
\{sattiraju,kochems,schotten\}@eit.uni-kl.de}}
\maketitle

\begin{abstract}
Short Range wireless devices are becoming more and more popular for ubiquitous sensor and actuator connectivity in industrial communication scenarios. Apart from communication-only scenarios, there are also mission-critical use cases where the distance between the two communicating nodes needs to be determined precisely. Applications such as Automatic Guided Vehicles (AGV's), Automatic Train Pairing additionally require the devices to scan the environment and detect any potential humans/obstacles. Ultra-Wide Band (UWB) has emerged as a promising candidate for Real-Time Ranging and Localization (RTRL) due to advantages such as large channel capacity, better co-existence with legacy systems due to low transmit power, better performance in multipath environments etc. In this paper, we evaluate the performance of a UWB COTS device - TimeDomain P440 which can operate as a ranging radio and a monostatic radar simultaneously. To this end, we evaluate the possibility of using Supervised Learning based estimators for predicting the presence of obstacles by constructing a multiclass hypothesis. Simulation results show that the Ensemble tree based methods are able to calculate the likelihood of obstacle collision with accuracies close to 95\%.
     
\end{abstract}

\section{Introduction}
{\let\thefootnote\relax\footnote{This is a preprint, the full paper is published in Proceedings of 13th International Workshop on Factory Communication Systems (IEEE WFCS 2017), \copyright 2017 IEEE. Personal use of this material is permitted. However, permission to use this material for any other purposes must be obtained from the IEEE by sending a request to pubs-permissions@ieee.org.}}

Real-Time Ranging \& Localization (RTRL) is an important requirement for \textit{Industry 4.0} scenarios such as Industrial Machine type communication, Vehicle-to-Vehicle (V2V) communications etc. that rely on the agent's relative location to each other. Some of the use cases additionally require that the two ranging entities include functionality to detect humans/obstacles and stop if necessary to avoid a collision. An example of such an use case is Automatic Train Pairing (ATP) for Train-Train communication where the two trains need to accurately and in real-time, measure their distances to each other in order to couple automatically. The moving train needs to periodically measure its range to the stationary train and simultaneously monitor the surroundings to detect potential humans/obstacles \cite{SBDist2016a}. Additionally, the software-wise pairing may also be initiated in the same time meaning that a communication channel according to TCN requirements\cite{IEC2005} be already established. An ideal wireless module hence would execute all the three operations - communications, ranging and obstacle detection simultaneously and in real-time.



However, from a design perspective, it is beneficial to distribute the above functions to independent modules working in tight real-time cohesion with each other. This also means ability to support existing state of the art COTS components since the increased demand for RTRL also saw the rise of technologies such as Visible Light Communication (VLC) and Ultra-Wide-Band (UWB) etc. System on Chips (SoC's) capable of accurate ranging such as TimeDomain P440 \cite{Petroff2012}, Decawave EVB1000 etc. have demonstrated high performance, utility and the path to lower costs. It is advantageous to integrate such existing Commercial Off the Shelf (COTS) components into one device. This can be done by making the device modular in the sense that each of the three functions can be realized by means of independently working modules that operate in tight cohesion with each other.  This is also the idea behind the proposed SBDist system\cite{SBDist2016b} that supports modularization by logically separating the modem functionality.  The ease of integration of such devices leads to development of hybrid systems where the traditional communication system is in charge of reliable data transfer whereas the integrated COTS components take care of functions such as ranging, obstacle detection etc and if necessary, also as a redundant/backup communication system.\

The aim of this paper is to demonstrate the utility of such an integrated communications system (P440 from TimeDomain) that performs realtime ranging and can simultaneously operate as a monostatic radar. The ranging performance of P440 has been investiagated in many prior works and was shown to achieve accuracies in the range of 2-10 cm\cite{Kasmi2017}. In this paper, we limit our investigation to obstacle detection using P440 as a monostatic radar. To this end, we collect the output radar scans (pulse responses) and use it as training data for multi-label classification by employing existing supervised machine learning algorithms.\

\section{Obstacle Detection using P440 as Monostatic Radar}


P440 captures the radar pulse response in a different way than conventional radars. Most radars will downconvert the radar pulse response, split the signal into an I and Q channel and then digitize the two baseband signals. In contrast, P440 will digitize the signal directly from the output of the antenna low noise amplifier (LNA) \cite{Petroff2012}. We can then use the Hilbert transform to produce the correponsing I and Q data streams which are then converted into frequency domain and further processed to calculate the range estimates of the signal. However, P440 does not have inbuilt mechanism for such radar processing and hence this has to be built externally in the software.\


\begin{figure}[h]
	\centering
	\includegraphics[width=0.48\textwidth]{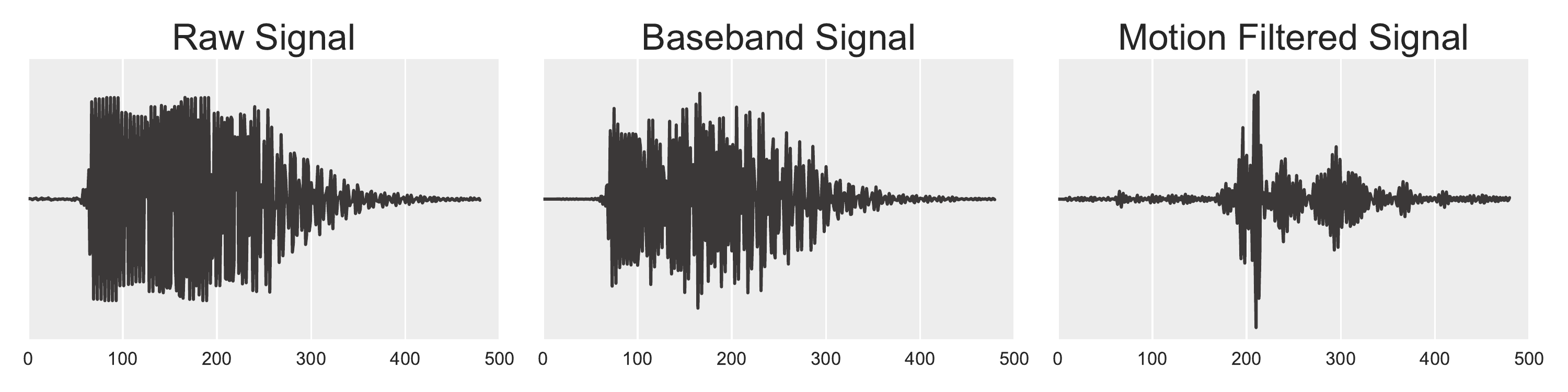}
	\caption{Received waveforms from P440}
	\label{tdsignals}
\end{figure} 

Taking advantage of the multipath resolution of the P440's radar, we used the motion filtered radar scans (using second order difference filter) in order to detect any obstacles. Figure~\ref{tdsignals} shows the the resulting P440's output waveforms for raw, baseband and motion filtered signals respectively. The horizontal axis indicates time in nanoseconds relative to the arrival time of the direct path and the vertical axis indicates response magnitude and polarity. It can be seen that the time of arrival of the direct path is clear for P440's pulsed UWB signal. This feature can be exploited to train a classifier and predict the presence of an obstacle by formulating a hypothesis as explained in the next section.\


\subsection{Machine Learning for Obstacle Detection}

Machine learning has found wide ranging applications in image/audio processing, finance, economics, social behavioral analysis, project management etc \cite{Alpaydin}. With the availability of big data, ML techniques are also being  increasingly applied in the field of wireless communications \cite{Jiang2016} \cite{Alsheikh2014} for finding optimal solutions in Antenna Selection \cite{Joung2016}, link adaptation\cite{DelTesta2016}, node detection\cite{DelTesta2016}\cite{Metcalf2015} and Wireless Security\cite{Weinand2017a}. ML techniques are also currently used in UWB based radars for sensitive applications such as gesture recognition \cite{Park2016} Sleep Apnea Screening \cite{Javaid2015} etc.

An ML algorithm learns the execution of a particular task $T$, maintaining a specific performance metric $M$, based on exploiting its experience $E$ \cite{Jiang2016}. ML algorithms can be classified into Supervised and Unsupervised learning, where supervised/unsupervised learning indicates the presence/absence of labeled samples in the input dataset.\

The supervised ML methods analyze the labeled training data and produce an inferred function, which can be used for mapping new examples. This requires the learning algorithm to generalize from the training data to unseen situations in a \textit{reasonable} way. Examples of supervised learning models include Linear models (Perceptron, Logistic Regression etc.), Support Vector Machines (SVMs), Nearest Neighbor methods, Tree based methods, Ensemble methods (Bagging and Boosting algorithms), Neural Networks (Multi-Layer Perceptron) etc.\

On the other hand, unsupervised ML methods infer a function to describe hidden structure from unlabeled data. Since the data is unlabeled, there is no objective evaluation for the accuracy of the algorithm's output. Examples of such methods include Gaussian Mixture models, clustering models (K-means, DBSCAN etc.), Principle Component Analysis, Neural Networks (Restricted Boltzmann machines) etc.\\

In this paper, we limit our evaluation to currently available supervised learning methods as outlined in Table~\ref{gridtable}. The details of each estimator can be found in \cite{Pedregosa2001}.

\begin{table*}[]
\centering
\caption{List of estimators and hyper-parameter grid}
\label{gridtable}
\begin{tabular}{|l|l|l|l|}
\hline
Name                                                                                                              & Underlying Model                                                  & Parameters                              & Values                                   \\ \hline
\multirow{2}{*}{Logistic Regression}                                                                              & \multirow{2}{*}{Linear}                                           & Regularization Parameter (C)            & {[}0.001, 0.01, 0.1, 1, 10, 100, 1000{]} \\ \cline{3-4} 
                                                                                                                  &                                                                   & Solver                                  & {[}lbfgs, sag, newton-cg{]}              \\ \hline
Perceptron                                                                                                        & Linear                                                            & Regularization Parameter (Alpha)        & {[}0.0001, 0.001, 0.01, 0.1, 1{]}        \\ \hline
K-Nearest neighbors                                                                                               & Nearest neighbor                                                  & Number of neighbors to consider (N)     & {[}1, 2, 3, ..., 30{]}                   \\ \hline
Linear SVC                                                                                                        & \begin{tabular}[c]{@{}l@{}}Support Vector\\ Machines\end{tabular} & Regularization Parameter (C)            & {[}0.001, 0.01, 0.1, 1, 10, 100, 1000{]} \\ \hline
\multirow{2}{*}{Decision Tree}                                                                                    & \multirow{2}{*}{Tree Based}                                       & Splitting Quality Measure               & {[}gini, entropy {]}                     \\ \cline{3-4} 
                                                                                                                  &                                                                   & Max\_features to consider for splitting & {[}auto, sqrt, log2 {]}                  \\ \hline
\multirow{3}{*}{\begin{tabular}[c]{@{}l@{}}Random Forest Classifier \\ and\\ Extra Trees Classifier\end{tabular}} & \multirow{3}{*}{Tree Based Ensemble}                              & Number of estimators (n)                & {[}16, 32, 64, 128, 256{]}               \\ \cline{3-4} 
                                                                                                                  &                                                                   & Splitting Quality Measure               & {[}gini, entropy {]}                     \\ \cline{3-4} 
                                                                                                                  &                                                                   & Max\_features to consider for splitting & {[}auto, sqrt, log2 {]}                  \\ \hline
\multirow{2}{*}{Gradient Boosting Classifier}                                                                     & \multirow{2}{*}{Tree Based}                                       & Number of estimators (n)                & {[}16, 32, 64, 128, 256 {]}              \\ \cline{3-4} 
                                                                                                                  &                                                                   & Learning Rate                           & {[}0.2, 0.5, 0.8, 1.0 {]}                \\ \hline
\end{tabular}
\end{table*}

\subsection{Construction of Hypothesis and Labeling}

The hypothesis of obstacle presence/absence can be constructed in many ways. We used the following labeling methods for construction of hypothesis and assigning labels to input radar data.\

%
%
%
%

\subsubsection{Simple Multi-Class Labeling}
\begin{figure}[h]
	\centering
	\begin{subfigure}[b]{.60\linewidth}
		\includegraphics[width=\linewidth]{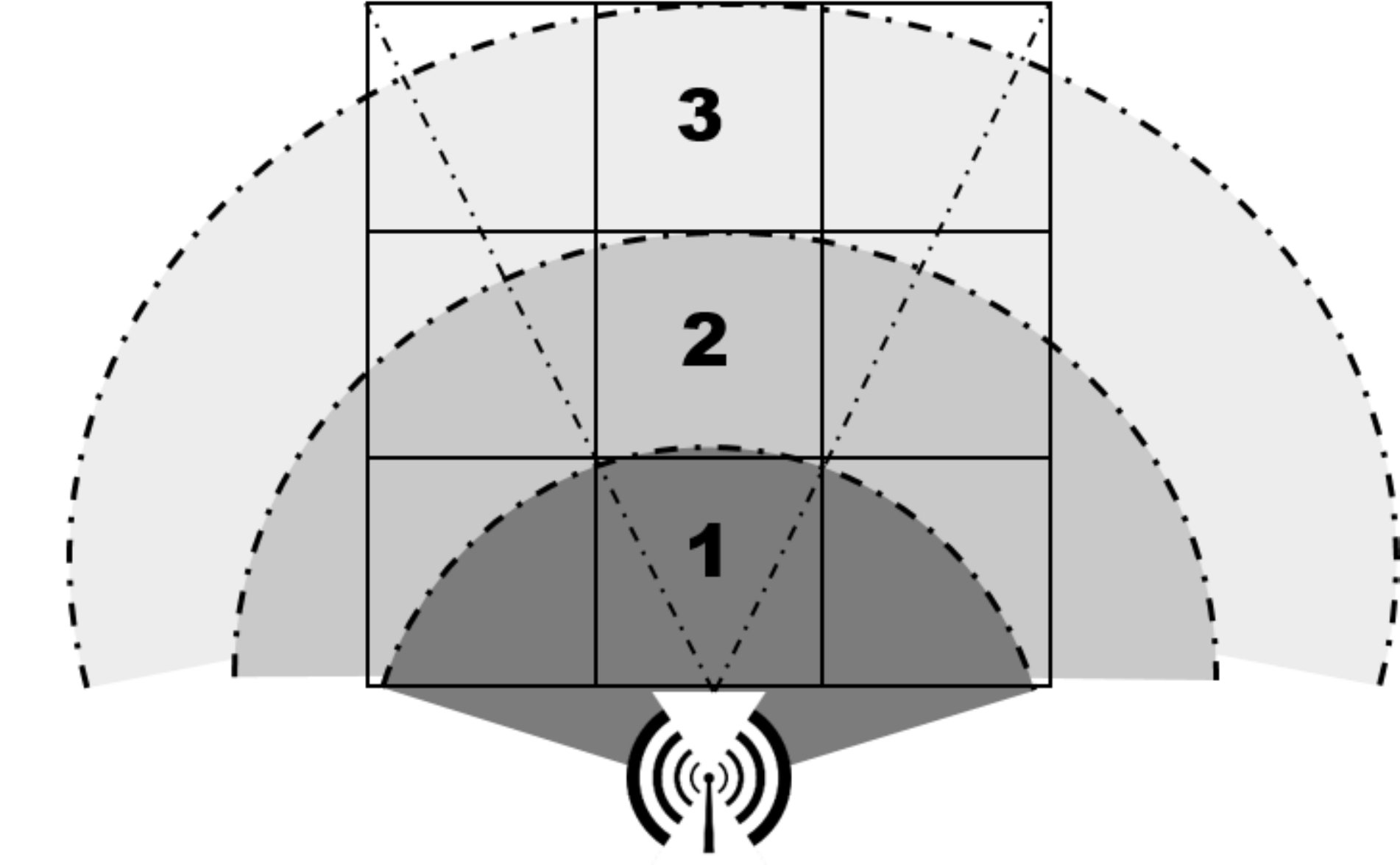}
		\caption{Simple Multi-Class Labeling}\label{simplelabeling}
	\end{subfigure}
	\begin{subfigure}[b]{.30\linewidth}
		\includegraphics[width=\linewidth]{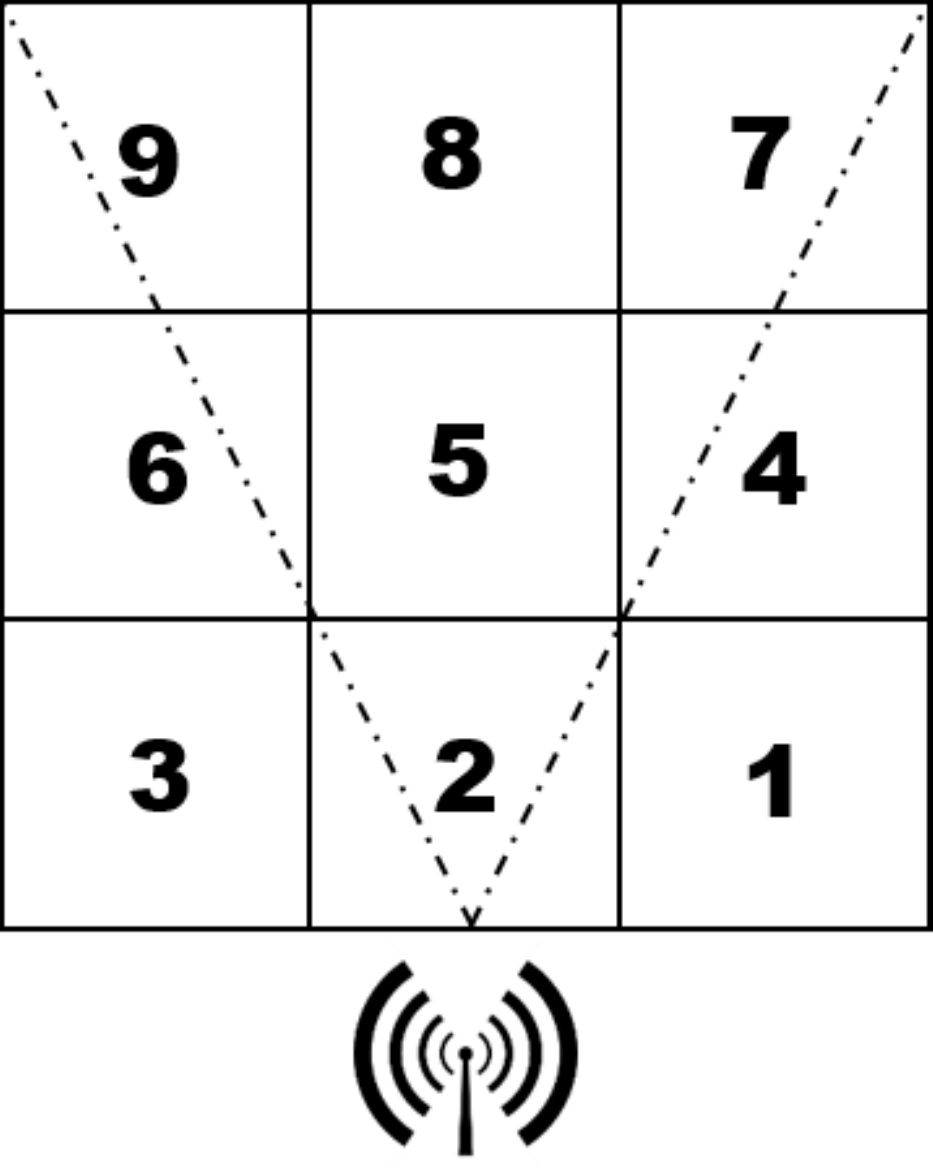}
		\caption{Grid Labeling}\label{gridlabeling}
	\end{subfigure}
	\caption{Labeling of data}
	\label{labeling}
\end{figure}

A binary hypothesis test for obstacle detection would output 0 when there is no person and 1 when there is a person. However, for the ATP use case, the radar must also be able to distinguish between a safely standing person/operator and a person in risk of collision. A well trained binary classifier would predict 1 in both these cases. Though both are correct predictions, we dont have any information of the obstacle's likelihood of collision. Hence, in this case, we can extend the binary hypothesis into a simple multi-class hypothesis as follows (Figure~\ref{simplelabeling})

\[ \text{H} = \left\{ 
\begin{array}{l l l l}
0, & \quad \text{If there is no person}\\
1, & \quad \text{High Risk person}\\
2, & \quad \text{Medium Risk person}\\
3, & \quad \text{Low Risk person}\\
\end{array} \right.\]

\subsubsection{Grid based Multi-Class Labeling}

We can also adapt a grid based approach for multi-class labeling as shown in Figure~\ref{gridlabeling}. In this case, each grid corresponds to an independent label for the training data. So for a $3X3$ grid, there are 10 labels including label $0$ for no person.

\[ \text{H} = \left\{ 
\begin{array}{l l l l}
0, & \quad \text{If there is no person}\\
1, & \quad \text{Person in Grid 1 }\\
.. & \quad \text{.... }\\
9, & \quad \text{Person in Grid 9 }\\
\end{array} \right.\]

\subsection{Evaluation Pipeline}
\begin{figure}[h]
	\centering
	\includegraphics[width=0.48\textwidth]{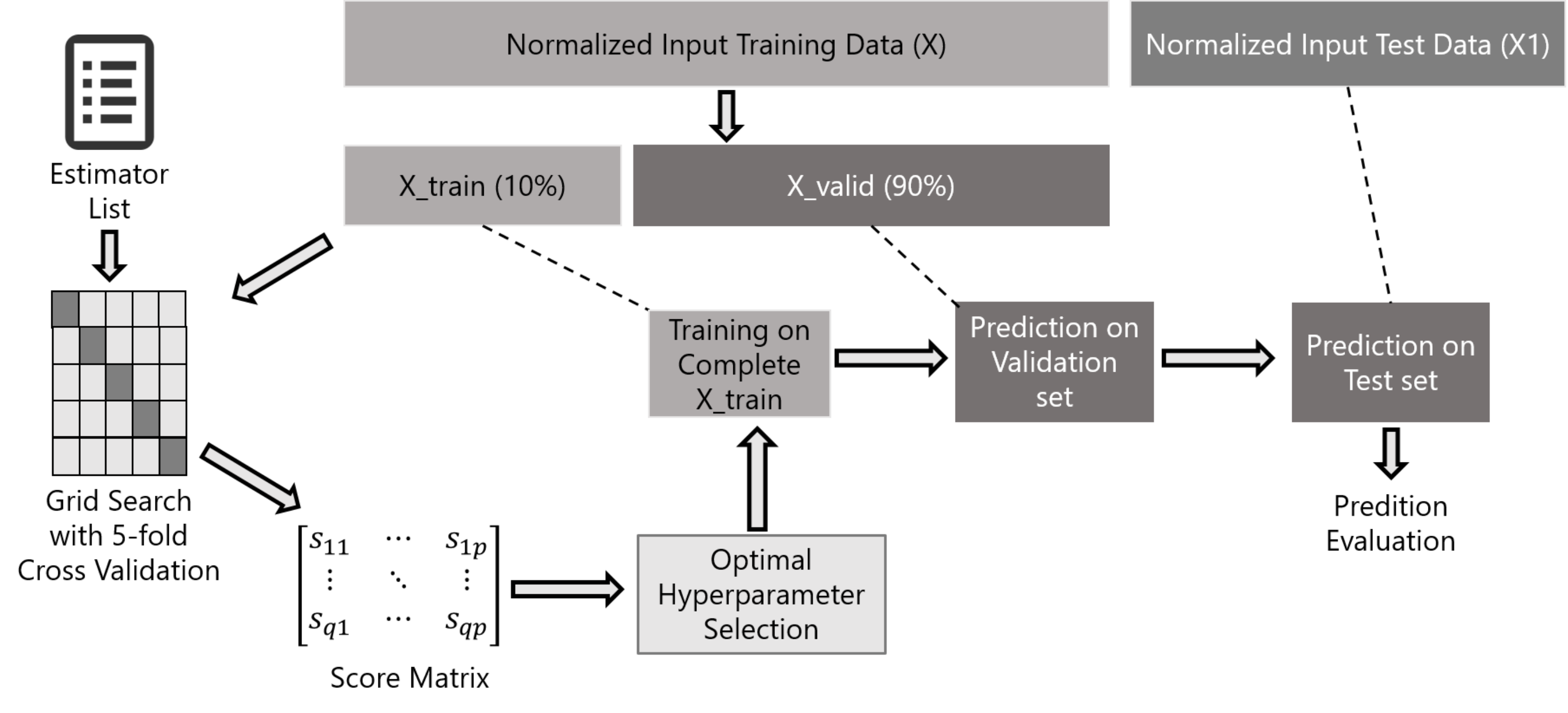}
	\caption{Simulation Pipeline}
	\label{pipeline}
\end{figure}

In order to evaluate the performance of the chosen ML classifiers, we used the pipeline as illustrated in Figure~\ref{pipeline} \

\subsubsection{Input Data Collection and Preparation}

For the purpose of generalization, we collected radar scans for 2 scenarios -  indoor building corridor and in the outdoor open area ($N_{sc} = 2$). For each scenario, we divided the area under interest radially (for 4-class hypopthesis) and into a $3X3$ grid for a 10-class hypothesis ($N_{ct} = 2$) and collected raw, baseband and motion filtered signals ($N_{dt}=3$). Hence the total number of data sets ($N_D$) is $N_{sc} * N_{dt} * N_{ct} = 12$. Each data set consists of independent \textit{training} and \textit{testing} data.\  

During offline evaluation, we used the motion filtered data set which was parsed sequentially to obtain the input data vectors for training data ($X$), training labels ($y$), testing data ($X_1$) and training labels ($y_1$). $X$ was further split into a training set ($X_{train}$ which is 10\% of the $X$) and a validation set ($X_{valid}$ with the remaining 90\%) as shown in Figure~\ref{pipeline}. All the data sets ($X_{train}$, $X_{valid}$ and $X_1$) were preprocessed by normalizing them around zero center as follows 

\begin{equation}
	 X_n = \sum_{i=1}^{N}\left (\frac{X_i-\frac{1}{N}\sum_{i=1}^{N}X_i}{\sqrt{\frac{1}{N}\sum_{i=1}^{N}(X_i-\mu)^{2}}}  \right )
\end{equation}

where $X_n$ is the normalized data, $X_i$ is the input data, $N$ is the length of input data, $\mu$ the mean and $\sigma$ the standard deviation.Preprocessing of data has advantages such as faster convergence of the Gradient Descent algorithm.\

%
%
%

\subsection{Model Evaluation \& Selection}

Hyper Parameters (HP's) are the tunable parameters of the estimator (e.g., regularization parameter for linear models, number of neighbors for knearest neighbors etc and their selection visibly effects the model's performance. In order to select the optimal parameters, we used exhaustive search over the grid space (Table~\ref{gridtable}) using $X_{train}$ and K-Fold cross validation. For each instantiation of the ML classifier with a given HP set, we divide $X_{train}$ into 5-folds (5 was chosen randomly) and preserving the ratio of labels in each fold. Hence, each training operation lasts 5 iterations where for each iteration, the model is tested on each fold by training on the other four folds. For the performance comparison (scoring), we used classification accuracy which is the number of correct predictions made divided by the total number of predictions made, multiplied by 100 to turn it into a percentage.\

\begin{figure}[h]
	\centering
	\includegraphics[width=0.48\textwidth]{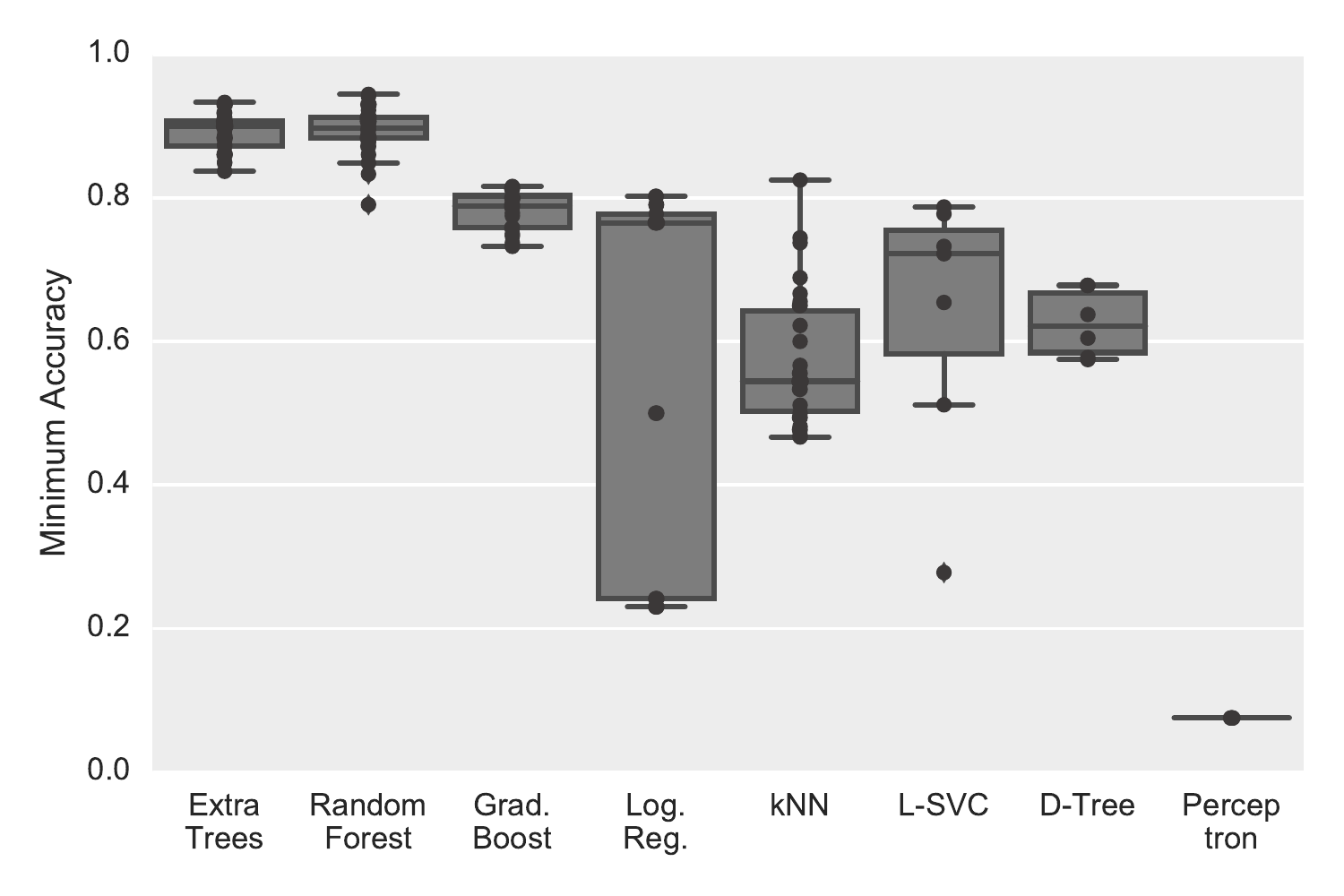}
	\caption{Scores of estimators during grid search}
	\label{scores}
\end{figure}

We used the following strategy to select the optimal hyperparameters. For every instantiation of the estimator, there are $k=5$ scores corresponding to each fold and as associated scores array $s$. We selected the minimum $s_{min}$ resulting in an array $(S_{min,1},S_{min,2},...,S_{min,n})$ where $n$ is the number of the estimator`s parameters in Table~\ref{gridtable}. The optimal estimator`s hyperparameters set is then selected by picking out the indices ($argmax$) of the maximum value $S_{max}$ from $S$.

\subsection{Model Fitting, Validation \& Testing}

The selected model is once again trained on the whole training data $X_{train}$ (since we omitted one fold per training operation during grid search). The trained model is then used for prediction  on the validation set ($X_{valid}$). Finally, the model is used to predict the labels on the test set $X_1$ so that we have the validation and testing performance for performance comparison.\

For ther performance comparison, we used the accuracy metric which can be defined as the percentage of correct predictions to the total number of predictions.\

\begin{figure}[h]
	\centering
	\begin{subfigure}[b]{\linewidth}
		\includegraphics[width=\linewidth]{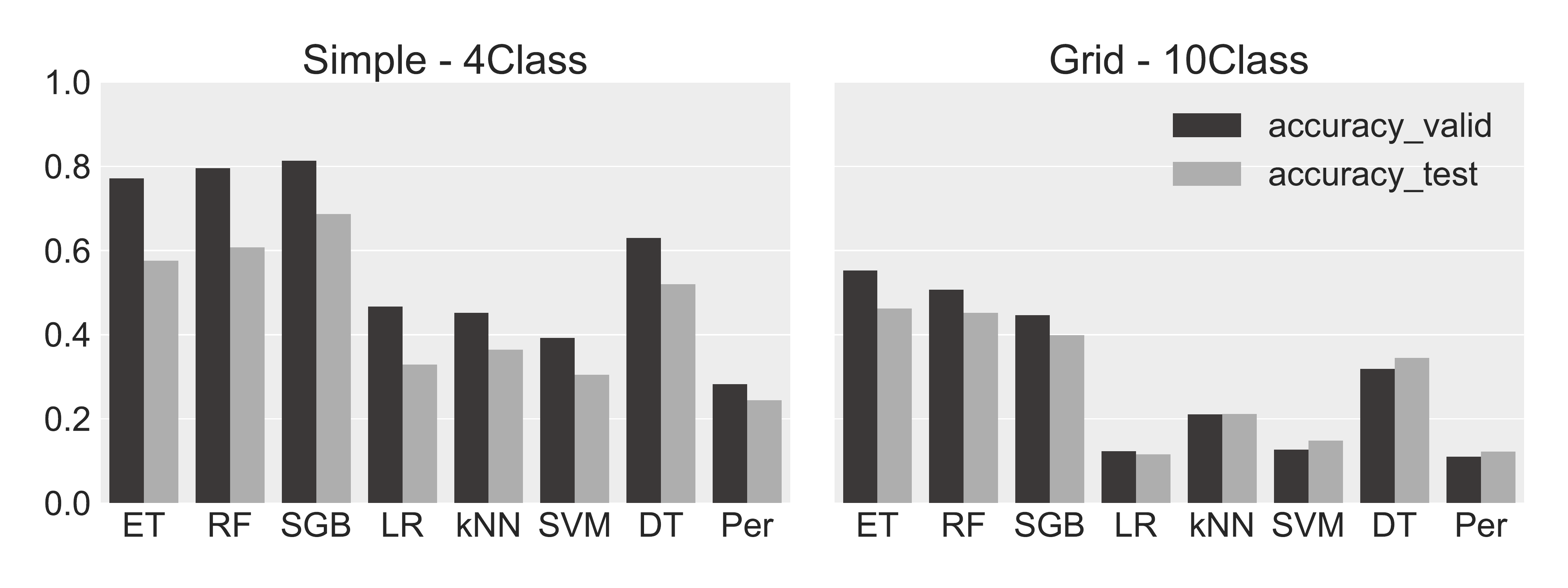}
		\caption{Indoor Scenario}\label{indoorscenario}
	\end{subfigure}\\
	\begin{subfigure}[b]{\linewidth}
		\includegraphics[width=\linewidth]{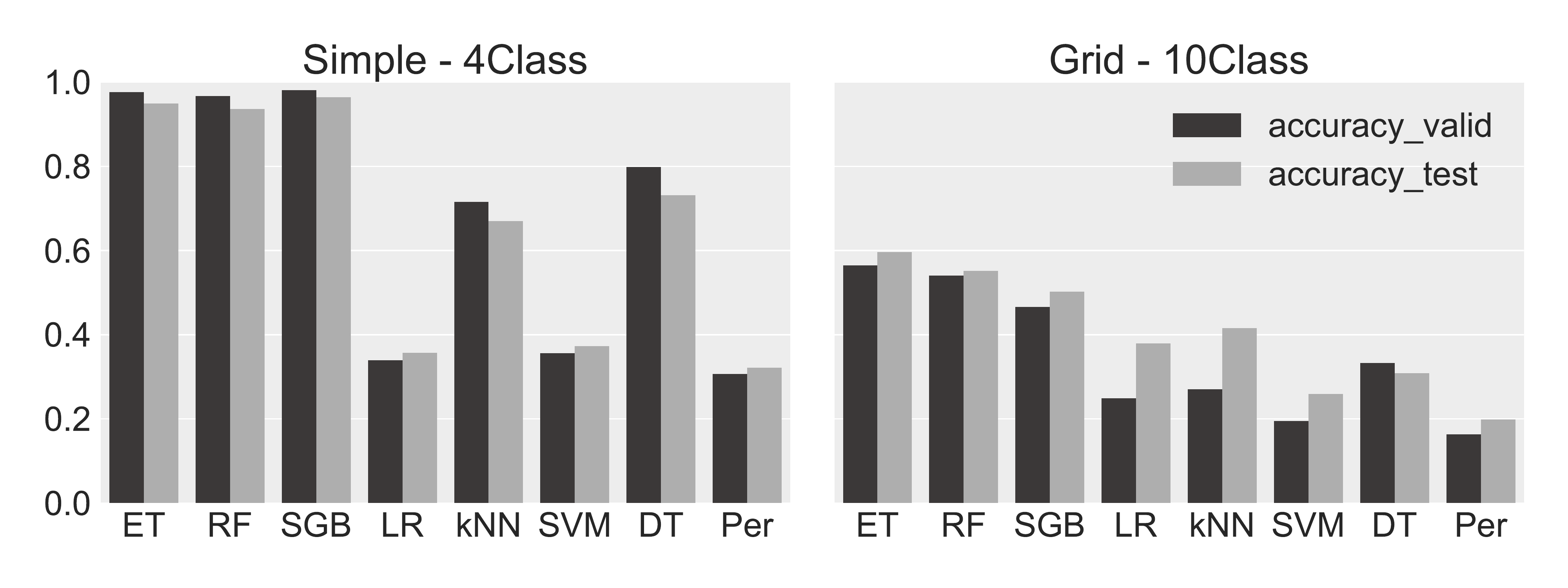}
		\caption{Outdoor Scenario}\label{outdoorscenario}
	\end{subfigure}
	\caption{Validation and Testing Accuracy}
	\label{accuracy}
\end{figure}


\subsection{Interpretation of Results}
Figures~\ref{accuracy} show the accuracies of the estimators (x-axis) for indoor (Figure~\ref{indoorscenario}) and outdoor (Figure~\ref{outdoorscenario}) scenarios respectively. In general, it can be seen that the tree based estimators (Decision Tree (DT), Extra Trees (ET) and Random Forests (RF)) along with the Stochastic Gradient Boost (SGB) exhibit better performance compared to linear models (Logistic Regression (LR), Perceptron (Per), Linear SVC (SVM)) and k-nearest neighbors (kNN). Evidently, the simple 4-class approach shows higher accuracies than grid based labeling. This is due to the ambiguity of the radar to distinguish similarly spaced targets that are on the left to those targets to the right.It can also be seen that the indoor scenario shows a relatively higher degradation of the validation and testing scores. This can be attributed to the rich multipath environment in the indoor scenario in which case the model fails to generalize well.\


Overall, for the simple 4-Class labeling, it can be seen that ET, RF and SGB show the best performance with accuracies of 55-70\% for indoor scenario and >95\% for outdoor scenario. This is a better performance than using the traditional tree based models (DT) alone that has an accuracy of 55\% and 70\% for indoor and outdoor scenarios respectively.\

\section{Conclusions and Future Work}
In order to realize the \textit{Industry 4.0} usecases, the  future wireless modules must be able to perform the functions of communication, ranging and obstacle detection simulteneously and in real-time. This paper discusses the approach of using existing  COTS components integrated tightly with the legacy communication system inorder to realize these functions. To this end, using the P440 module and its radar scans, we were able to construct a simple and a grid based multi-class hypothesis to test the presence/absence of an obstacle using Supervised Machine Learning algorithms. The accompanying simulation results show that the tree based ensemble methods such as Random Forest (RF) and Extra Tree (ET) classifiers perform well with classification accuracies of >95\% and with faster training and testing times. However, our models are still susceptible to noise which can be mitigated by the availability of big data sets. Moreover, we only deal with supervised learning algorithms in this paper which leaves the evaluation of Unsupervised learning algorithms and neural networks in the future.\

\bibliographystyle{IEEEtran}
\bibliography{Ranging_ML}

\section*{Acknowledgements}
Part of this work has been performed in the framework of the BMBF project SBDist. The authors would like to acknowledge the contributions of their colleagues, although the views expressed are those of the authors and do not necessarily represent the project.
\end{document}